\documentclass[superscriptaddress,pre,showpacs,preprintnumbers,amsmath,amssymb,aps]{revtex4}

\usepackage{graphicx}
\usepackage{epstopdf}
\usepackage{dcolumn}
\usepackage{bm}
\usepackage{color}

\newcommand{\fst}{f^{*}}

\newcommand{\cs}{c_{\rm s}}

\newcommand{\integral}[1]{\left\langle{#1}\right\rangle}

\begin{document}

\preprint{Submitted to Physical Review E, 11-MAR-2011, revised}

\title{Enhanced lattice Bhatnagar-Gross-Krook method \\
for fluid dynamics simulation}

\author{Ilya V. Karlin}\email{karlin@lav.mavt.ethz.ch}
\affiliation {Energy Technology Research Group, School of Engineering Sciences, University of Southampton, Southampton, SO17 1BJ, UK}
\affiliation{Aerothermochemistry and Combustion Systems Lab, ETH Zurich, 8092 Zurich, Switzerland}

\author{Daniel Lycett-Brown}\email{djlb1e08@soton.ac.uk}
\affiliation {Energy Technology Research Group, School of Engineering Sciences, University of Southampton, Southampton, SO17 1BJ, UK}

\author{Kai H. Luo}\email{k.h.luo@soton.ac.uk}
\affiliation {Energy Technology Research Group, School of Engineering Sciences, University of Southampton, Southampton, SO17 1BJ, UK}

\date{\today}

\begin{abstract}
A generalization of the lattice Bhatnagar-Gross-Krook (LBGK) model for the simulation of hydrodynamics is presented, which
takes into account the difference and the frame-independence of the relaxation of non-hydrodynamic modes.
The present model retains the computationally efficient standard LBGK form with the generalized equilibrium explicitly derived.
The two-dimensional realization on the standard lattice is discussed in detail.
Performance of the model is assessed through a shear layer simulation and enhanced stability and accuracy with respect to the standard LBGK are reported. The results demonstrate that the present model is a useful upgrade of the standard LBGK without compromising its computational efficiency and accuracy.
\end{abstract}

\maketitle

\section{Introduction}
\label{sec:intro}

The lattice Bhatnagar-Gross-Krook (LBGK) model was conceived about twenty years ago \cite[]{Chen92,Qian92} and has rapidly taken a dominant role in the lattice Boltzmann approach to the simulation of complex hydrodynamic phenomena \cite[]{SucciRev,Aidun10}.
The success of LBGK is primarily based on its computational efficiency, accuracy and stability at moderate Reynolds number simulations.
However the standard LBGK has its limits in addressing direct numerical simulation of high Reynolds number flows caused by severe numerical instabilities triggered at a sub-grid scale whenever the grid is coarsened. This prompted a number of studies aimed at improving the LBGK method, among which we mention the unconditionally stable entropic LBGK model \cite{Karlin99}, a family of the matrix lattice Boltzmann models \cite[]{HSB,dHumieres92,Ladd94,Ladd94a,Geier2006,Shan2007}, the recent multi-step kinetic models \cite{Asinari09,Asinari10,Karlin2011}, and kinematically complete LBGK models on higher-order lattices \cite{Chikatamarla2010}.
While enhancing the stability of the standard LBGK model, the above approaches also have to answer questions about accuracy and computational efficiency.
In particular, the idea of using separate relaxation times for various non-hydrodynamic modes can be realized in various ways, and some of the realizations
may severely affect the accuracy of the simulation (see, e. g. \cite{Dellar2003,Freitas2011} and references therein).
At present, the mainstream of lattice Boltzmann research
remains with the LBGK, due to its unsurpassed simplicity, computational efficiency and acceptable accuracy.

Under such a state of affairs it appears that an enhancement of the standard LBGK model with respect to stability, but without a compromise on
computational simplicity and accuracy, is needed. In this work, we aim at precisely this kind of enhancement of the standard LBGK.
Below we shall introduce a lattice Boltzmann model taking into account different relaxation rates for various non-hydrodynamic modes in a co-moving reference frame. This model assumes a LBGK form with a generalized equilibrium explicitly obtained (here, for the two-dimensional case) and it does not incur any significant computational overhead with respect to the standard LBGK. Extensive simulations of a chosen benchmark flow (roll up of a shear layer) reveal that the enhanced LBGK model features significantly increased stability, while retaining the accuracy of the LBGK. On the practical side, the present numerical algorithm is simple, requiring just a few line changes in existing standard LBGK codes.

The outline of the paper is as follows: In section \ref{sec:EnhLBGK}, details of the construction of the enhanced LBGK model in two dimensions are presented. In section \ref{sec:results}, the stability and accuracy of the present scheme is assessed in a benchmark simulation of shear layer vortical flow. Finally, some conclusions are drawn in section \ref{sec:conclu}.

\section{Enhanced LBGK model in two dimensions}
\label{sec:EnhLBGK}

\subsection{Moment representation}

For the sake of presentation and without any loss of generality, we consider the popular nine-velocity model, the so-called D2Q9 lattice. The discrete velocities are constructed as a tensor product of two one-dimensional velocity sets, $v_{(i)}=i$, where $i=0,\pm 1$; thus $v_{(i,j)}=(v_{(i)},v_{(j)})$ in the fixed Cartesian reference frame. Populations are labeled accordingly, $f_{(i,j)}$.
We start with the moment representation of the populations. To this end we recall that any product lattice, such as the D2Q9, is characterized  by  natural moments (cf.\ e.g. \cite{Karlin10}). For D2Q9, these natural moments are $\rho M_{pq}$, where $\rho=\integral{f_{(i,j)}}$ is the density, and
\begin{equation}
\label{eq:natmoments}
\rho M_{pq}=\langle f_{(i,j)}v_{(i)}^p v_{(j)}^q\rangle,\ p,q\in\{0,1,2\}.
\end{equation}
Notation $\integral{...}$ is used as a shorthand for summation over all the velocity indices as displayed.
In the sequel, we use the following linear combinations to represent natural moments (\ref{eq:natmoments})
\begin{equation}\label{eq:natmoments1}
M_{00}, u_x=M_{10},\ u_y=M_{01},\ T=M_{20}+M_{02},\ N=M_{20}-M_{02},\ \Pi_{xy}=M_{11},\ Q_{xyy}=M_{12},\ Q_{yxx}=M_{21},\ A=M_{22}.
\end{equation}
For the sake of completeness we recall that these are usually interpreted as the normalization to the density ($M_{00}=1$), the flow velocity components ($u_x$, $u_y$), the trace of the pressure tensor at unit density ($T$), the normal stress difference at unit density ($N$), and the  off-diagonal component of the pressure tensor at unit density ($\Pi_{xy}$). The (linearly independent) third-order moments ($Q_{xyy}$, $Q_{yxx}$) and the fourth-order moment ($A$) lack a direct physical interpretation. However, as we shall see below, they are of special importance for achieving better performance of the lattice Boltzmann schemes.

With the set of natural moments (\ref{eq:natmoments1}), populations are uniquely represented as follows ($\sigma,\lambda=\{-1,1\}$):
\begin{align}
\begin{split}\label{eq:momentsD2Q9}
f_{(0,0)}&=\rho\left(1-T+A\right),\\
f_{(\sigma,0)}&=\frac{1}{2}\rho\left(\frac{1}{2}(T+N)+\sigma u_x-\sigma Q_{xyy}-A\right),\\
{f}_{(0,\lambda)}&=\frac{1}{2}\rho\left(\frac{1}{2}(T-N)+\lambda u_y-\lambda Q_{yxx}-A\right),\\
f_{(\sigma,\lambda)}&=\frac{1}{4}\rho\left(A+(\sigma)(\lambda)\Pi_{xy}+\sigma Q_{xyy}+\lambda Q_{yxx}\right).
\end{split}
\end{align}
We note in passing that a subset of the populations (\ref{eq:momentsD2Q9}) specified by the closure relations, $A=M_{20}M_{02}$, $Q_{xyy}=M_{10}M_{02}$ and $Q_{yxx}=M_{01}M_{20}$ gives an example of a fully factorized population termed unidirectional quasi-equilibrium (UniQuE) in \cite{Karlin10}, and it is used as an intermediate quasi-equilibrium in some constructions \cite{Asinari09,Asinari10}. We shall use a different route here.

In order to construct an analog of the standard LBGK model, we further introduce central moments of the form
\begin{equation}
\rho\tilde{M}_{pq}=\integral{(v_{(i)}-u_x)^p(v_{(j)}-u_y)^qf},
\end{equation}
and use identity
\begin{align}\label{eq:cm}
\begin{split}
&\Pi_{xy}=\tilde{\Pi}_{xy}+u_xu_y,\\
&N=\tilde{N}+(u_x^2-u_y^2),\\
&T=\tilde{T}+u^2,\\
&Q_{xyy}=\tilde{Q}_{xyy}+2u_y\tilde{\Pi}_{xy}-\frac{1}{2}u_x\tilde{N}+\frac{1}{2}u_x\tilde{T}+u_x u_y^2,\\
&Q_{yxx}=\tilde{Q}_{yxx}+2u_x\tilde{\Pi}_{xy}+\frac{1}{2}u_y\tilde{N}+\frac{1}{2}u_y\tilde{T}+u_y u_x^2,\\
&A=\tilde{A} +2\left[u_x\tilde{Q}_{xyy}+u_y\tilde{Q}_{yxx}\right]
+4u_xu_y\tilde{\Pi}_{xy}+\frac{1}{2}u^2 \tilde{T}-\frac{1}{2}(u_x^2-u_y^2)\tilde{N}+u_x^2u_y^2.\\
\end{split}
\end{align}
We remark in passing that the mapping of natural moments onto central moments is nonlinear (it explicitly depends on the powers of the velocity components).
Therefore, implementation of their relaxation in the framework of the matrix model with a fixed transformation matrix from moments to populations becomes involved,
which may negatively affect efficiency and accuracy \cite{Geier2006}.

Using the central moments representation, Eq. (\ref{eq:momentsD2Q9}) is rewritten upon substituting (\ref{eq:cm}) into (\ref{eq:momentsD2Q9}) and rearranging terms,
\begin{align}
\begin{split}\label{eq:momentsD2Q9c}
f_{(0,0)}&=\rho\left(1 +u_x^2u_y^2-u^2\right)\\
&+\rho\left(4u_xu_y\tilde{\Pi}_{xy} -\left[\frac{u_x^2-u_y^2}{2}\right]\tilde{N}\right)
+ \rho\left(\left[\frac{u^2-2}{2}\right] \tilde{T} +2u_x\tilde{Q}_{xyy}+2u_y\tilde{Q}_{yxx}+ \tilde{A}\right),\\
f_{(\sigma,0)}&= \frac{\rho}{2}(u_x^2 +\sigma u_x(1-u_y^2)- u_x^2u_y^2)\\
&+\frac{\rho}{2}\left(\left[\frac{1+\sigma u_x + u_x^2-u_y^2}{2}\right] \tilde{N} -(2\sigma u_y+4u_xu_y)\tilde{\Pi}_{xy}\right)\\
&+\frac{\rho}{2}\left(\left[\frac{1-\sigma u_x-u^2}{2}\right] \tilde{T}-(\sigma + 2 u_x)\tilde{Q}_{xyy} -2 u_y\tilde{Q}_{yxx}-\tilde{A} \right),\\
f_{(0,\lambda)}&= \frac{\rho}{2}(u_y^2 +\lambda u_y(1-u_x^2)- u_x^2u_y^2)\\
&+\frac{\rho}{2}\left(\left[\frac{-1-\lambda u_y+u_x^2-u_y^2}{2}\right] \tilde{N} -(2\lambda u_y+4u_xu_y)\tilde{\Pi}_{xy}\right)\\
&+\frac{\rho}{2}\left(\left[\frac{1-\lambda u_y-u^2}{2}\right] \tilde{T}-(\lambda + 2 u_y)\tilde{Q}_{yxx} -2 u_x\tilde{Q}_{xyy}-\tilde{A} \right),\\
f_{(\sigma,\lambda)}&=\frac{\rho}{4}(\sigma \lambda u_xu_y+\sigma u_x u_y^2 + \lambda u_y u_x^2 +u_x^2u_y^2)\\
&+\frac{\rho}{4}\left( \left(4u_xu_y +(\sigma)(\lambda) +2\sigma u_y + 2\lambda u_x\right)\tilde{\Pi}_{xy}+\left[\frac{-u_x^2+u_y^2 -\sigma u_x +\lambda u_y}{2}\right] \tilde{N}\right)\\
&+\frac{\rho}{4}\left(\left[\frac{u^2 +\sigma u_x+\lambda u_y}{2}\right] \tilde{T} +(\sigma+2u_x)\tilde{Q}_{xyy}+(\lambda+2u_y)\tilde{Q}_{yxx}
+\tilde{A}\right).\\
\end{split}
\end{align}
Thus, any population on the D2Q9 lattice is uniquely represented by a linear combination of higher-order central moments with the coefficient of the linear combination being nonlinear functions of the flow velocity. While quite straightforward, this representation is helpful for the next steps of the construction.

\subsection{Enhanced LBGK scheme}

At the equilibrium, the higher-order (central) moments assume the following values (as dictated by the Maxwell-Boltzmann distribution, cf. e.g. \cite{Karlin10}):
\begin{align}\label{eq:eqmomc}
\begin{split}
\tilde{\Pi}_{xy}^{\rm eq}=\tilde{N}^{\rm eq}=\tilde{Q}_{xyy}^{\rm eq}=\tilde{Q}_{yxx}^{\rm eq}=0,\
\tilde{T}^{\rm eq}=2\cs^2,\
\tilde{A}^{\rm eq}=\cs^4,\\
\end{split}
\end{align}
where $\cs^2$ is the speed of sound squared (reference temperature) of the D2Q9 lattice,
\begin{equation}
\label{eq:cs}
\cs^2=\frac{1}{3}.
\end{equation}
Let us introduce four relaxation parameters,
\begin{equation}
\omega,\ \omega_b,\ \omega_3,\ \omega_4,
\end{equation}
and three ratios,
\begin{equation}\label{eq:relaxratios}
r_b=\frac{\omega_b}{\omega},\ r_3=\frac{\omega_3}{\omega},\ r_4=\frac{\omega_4}{\omega}.
\end{equation}
We consider a four-parametric family of lattice kinetic equations, written in the LBGK form (i.e. diagonal in the population representation),
\begin{align}
\begin{split}\label{eq:QEmodel}
f(\bm{x}+\bm{v},t+1)-f(\bm{x},t)=-\omega(f-\fst),
\end{split}
\end{align}
where the function $\fst$ (generalized equilibrium) is constructed as follows:
\begin{itemize}
\item For any higher-order central moment $\tilde{M}$, introduce a line segment connecting the current value $\tilde{M}$ with the equilibrium value thereof, $\tilde{M}^{\rm eq}$. This linear function will be parameterized with the parameter $r$, and denoted as $L_{r}[\tilde{M},\tilde{M}^{\rm eq}]$:
    \begin{equation}\label{eq:lincomb}
    L_{r}[\tilde{M},\tilde{M}^{\rm eq}]=(1-r)\tilde{M}+r\tilde{M}^{\rm eq}.
    \end{equation}
\item In the central moment representation, Eq.\ (\ref{eq:momentsD2Q9c}), replace the second-order moments $\tilde{\Pi}_{xy}$ and $\tilde{N}$ (responsible for the shear) by their values at the equilibrium, and replace the rest of the higher-order moments ($\tilde{T}$, responsible for the compressibility, and the third- and fourth-order moments, $\tilde{Q}_{xyy}$, $\tilde{Q}_{yxx}$, and $\tilde{A}$, respectively) by the linear combinations (\ref{eq:lincomb}) with the parametrization according to the ratios of relaxation rates (\ref{eq:relaxratios}). That is, if the shorthand notation is used for Eq.\ (\ref{eq:momentsD2Q9c}),
    \[f=f(\rho, \bm{u}, \tilde{\Pi}_{xy}, \tilde{N}, \tilde{T}, \tilde{Q}_{xyy}, \tilde{Q}_{yxx}, \tilde{A}),\]
      we set the generalized equilibrium in the LBGK-like kinetic equation (\ref{eq:QEmodel}) as follows:
\begin{equation}\label{eq:GenEQ}
\fst=f(\rho, \bm{u}, \tilde{\Pi}_{xy}^{\rm eq}, \tilde{N}^{\rm eq}, L_{r_b}[\tilde{T},\tilde{T}^{\rm eq}], L_{r_3}[\tilde{Q}_{xyy},\tilde{Q}^{\rm eq}_{xyy}],  L_{r_3}[\tilde{Q}_{yxx},\tilde{Q}^{\rm eq}_{yxx}], L_{r_4}[\tilde{A},\tilde{A}^{\rm eq}]).
\end{equation}
\end{itemize}

In the expanded form, the generalized equilibrium (\ref{eq:GenEQ}) reads:
\begin{align}
\begin{split}\label{eq:momentsD2Q9st}
\fst_{(0,0)}&=\rho\left(1 +u_x^2u_y^2-u^2\right)+ \rho\left(\frac{u^2-2}{2}\right) [(1-r_b)\tilde{T}+2r_b \cs^2] \\ &+\rho\left((1-r_3)[2u_x\tilde{Q}_{xyy}+2u_y\tilde{Q}_{yxx}]
+ [(1-r_4)\tilde{A}+r_4 \cs^4]\right),\\
\fst_{(\sigma,0)}&= \frac{\rho}{2}(u_x^2 +\sigma u_x(1-u_y^2)- u_x^2u_y^2)
+\rho\left(\frac{1-\sigma u_x-u^2}{4}\right)[(1-r_b)\tilde{T}+2r_b \cs^2]\\
&-\frac{\rho}{2}\left((1-r_3)[(\sigma + 2 u_x)\tilde{Q}_{xyy}
+2 u_y\tilde{Q}_{yxx}]+[(1-r_4)\tilde{A}+r_4 \cs^4]\right),\\
\fst_{(0,\lambda)}&= \frac{\rho}{2}(u_y^2 +\lambda u_y(1-u_x^2)- u_x^2u_y^2)
+\rho\left(\frac{1-\lambda u_y-u^2}{4}\right)[(1-r_b)\tilde{T}+2r_b \cs^2]\\
&-\frac{\rho}{2}\left((1-r_3)[(\lambda + 2 u_y)\tilde{Q}_{yxx} + 2 u_x\tilde{Q}_{xyy}]+[(1-r_4)\tilde{A}+r_4 \cs^4] \right),\\
\fst_{(\sigma,\lambda)}&=\frac{\rho}{4}(\sigma \lambda u_xu_y+\sigma u_x u_y^2 + \lambda u_y u_x^2 +u_x^2u_y^2)
+\rho\left(\frac{u^2 +\sigma u_x+\lambda u_y}{8}\right)[(1-r_b)\tilde{T}+2r_b \cs^2]\\
&+\frac{\rho}{4}\left((1-r_3)[(\sigma+2u_x)\tilde{Q}_{xyy}+(\lambda+2u_y)\tilde{Q}_{yxx}]
+[(1-r_4)\tilde{A}+r_4 \cs^4]\right).\\
\end{split}
\end{align}
Here we have taken into account the actual values of the higher-order moments at the equilibrium (\ref{eq:eqmomc}).
Using the standard multi-scale (Chapman-Enskog) analysis, it can be shown that (\ref{eq:QEmodel}) recovers the isothermal Navier-Stokes equations at reference temperature $T_0=\cs^2$,

\begin{align}
\begin{split}
&\partial_t \rho +\partial_{\alpha}(\rho u_{\alpha})=0,\\
&\partial_t u_{\alpha}+u_{\beta}\partial_{\beta} u_{\alpha}+\frac{1}{\rho}\partial_{\alpha}(\cs^2 \rho)
-\frac{1}{\rho}
\partial_{\beta}\left[\nu\rho\left(\partial_{\alpha}u_{\beta}+\partial_{\beta}u_{\alpha}-
\frac{2}{D}\delta_{\alpha\beta}\partial_{\gamma}u_{\gamma}\right)\right]-\frac{2}{D\rho}\partial_{\alpha}\left(\xi\rho\partial_{\gamma}u_{\gamma}\right)=0,
\end{split}
\end{align}
where $D=2$ is the spatial dimension, and
where the two viscosity coefficients, the kinematic (shear) viscosity $\nu$ and the bulk viscosity $\xi$ are
\begin{equation}
\nu=\left(\frac{1}{\omega}-\frac{1}{2}\right)\cs^2,\ \xi=\left(\frac{1}{\omega_b}-\frac{1}{2}\right)\cs^2.
\end{equation}

Below, we shall consider the case $\omega=\omega_b$ ($r_b=1$) (shear and bulk viscosities equal). Moreover, Eq.\ (\ref{eq:momentsD2Q9st}) can be further simplified by neglecting all the terms of $\mathcal{O}(u^3)$ and higher:
\begin{align}
\begin{split}\label{eq:momentsD2Q9stsimpl}
\fst_{(0,0)}&=\rho\left\{1- 2\cs^2 - (1-\cs^2) u^2+ (1-r_3)[2u_x\tilde{Q}_{xyy}+2u_y\tilde{Q}_{yxx}]
+ [(1-r_4)\tilde{A}+r_4 \cs^4]\right\},\\
\fst_{(\sigma,0)}&= \frac{\rho}{2}\left\{(1-\cs^2)\sigma u_x +u_x^2+(1-u^2)\cs^2
-
(1-r_3)[(\sigma + 2 u_x)\tilde{Q}_{xyy}
+2u_y\tilde{Q}_{yxx}]-[(1-r_4)\tilde{A}+r_4 \cs^4]\right\},\\
\fst_{(0,\lambda)}&= \frac{\rho}{2}\left\{(1-\cs^2)\lambda u_y +u_y^2+(1-u^2)\cs^2
- (1-r_3)[(\lambda + 2 u_y)\tilde{Q}_{yxx} +2 u_x\tilde{Q}_{xyy}]-[(1-r_4)\tilde{A}+r_4 \cs^4] \right\},\\
\fst_{(\sigma,\lambda)}&=\frac{\rho}{4}\left\{(\sigma u_x+\lambda u_y)\cs^2 +\sigma \lambda u_xu_y +u^2\cs^2
+ (1-r_3)[(\sigma+2u_x)\tilde{Q}_{xyy}+(\lambda+2u_y)\tilde{Q}_{yxx}]
+[(1-r_4)\tilde{A}+r_4 \cs^4]\right\}.\\
\end{split}
\end{align}
It can be readily checked that, by setting $r_3=r_4=1$, and using the value of the speed of sound as given by Eq.\ (\ref{eq:cs}), the generalized equilibrium (\ref{eq:momentsD2Q9stsimpl}) becomes the standard second-order equilibrium of the LBGK model, whereas at $r_b=r_3=r_4=1$, function (\ref{eq:momentsD2Q9st}) becomes the Maxwell equilibrium on this product-lattice (cf.\ \cite{Karlin10}).

In summary, the four-parametric enhanced LBGK model is fully explicit, and is defined by the four parameters ($\omega$, and the ratios $r_b$, $r_3$ and $r_4$) in the generalized equilibrium populations $\fst$ (\ref{eq:momentsD2Q9st}). Note that this is the maximal parametrization which correctly takes into account symmetries of the moments. The proposed model
is readily implemented, similar to the standard LBGK itself. Below we shall demonstrate the gain of the present model [with the restricted set of parameters, Eq.\ (\ref{eq:momentsD2Q9stsimpl})] with respect to the standard LBGK by considering a benchmark simulation of a shear flow.

\section{Results}
\label{sec:results}

\subsection{Stability}

We shall first assess the stability of the present scheme with respect to the standard LBGK. For this purpose a perturbed double periodic shear layer flow is used, with initial conditions
\begin{equation}
\label{eq:InitCon}
\begin{array}{ll}
u_x = \left\{ \begin{array}{cc}
                             U \tanh\left(\kappa \left(\frac{y}{L}-\frac{1}{4}\right) \right), y \leq \frac{L}{2}, \\
                             U \tanh\left(\kappa \left(\frac{3}{4}-\frac{y}{L}\right) \right), y > \frac{L}{2},
                           \end{array}\right. \\
u_y = \delta \sin\left(2\pi \left(x + \frac{1}{4} \right) \right),
\end{array}
\end{equation}
as studied by Minion and Brown \cite{Minion97}. $L$ is the number of grid points in both $x$ and $y$ directions, and periodic boundary conditions are applied in both directions. Varying the parameter $\kappa$ alters the width of the shear layers, and this is fixed at $\kappa$ = 80 throughout the following. The velocity perturbation in the $y$-direction initiates a Kelvin-Helmholtz instability causing the roll up of the anti-parallel shear layers. The parameter $\delta$ controls the size of the initial perturbation and is fixed here at $\delta$ = 0.05. $U$ determines the magnitude of the initial $x$-velocity.

Stability regimes for the parameters $r_3$ and $r_4$ are considered at a fixed Reynolds number, and then the stability limits of the Reynolds number are considered by independently varying $r_3$ and $r_4$. For this the Reynolds number is defined as
\begin{equation}
\label{eq:Re}
Re = \frac{3UL}{\frac{1}{\omega} - \frac{1}{2}}.
\end{equation}
$U$ is held constant throughout, with Reynolds number (at a fixed $L$) being varied by $\omega$ alone. As $U$ is constant, the initial Mach number, given by
\begin{equation}
\label{eq:Ma}
Ma = U \sqrt{3},
\end{equation}
is the same for each simulation. $U = 0.04$ is used, giving $Ma \approx 0.07$.

To determine stability, simulations were run for a large number of time steps, $T$, with instability being determined by any deviation in the total mass inside the domain. These time steps correspond to a time given by
\begin{equation}
\label{eq:Time}
t = \frac{TU}{L}.
\end{equation}
Here $T = 200,000$ was used, giving $t = 62.5$.

\begin{figure}[ht]
  \begin{center}
    \includegraphics[width=0.5\textwidth]{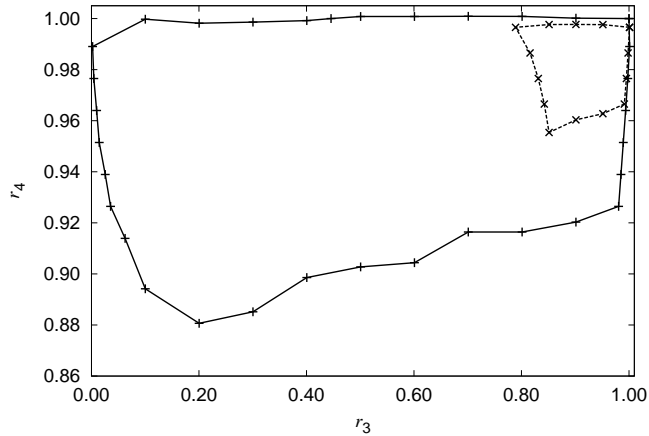}
  \end{center}
  \caption{$r_3$ vs $r_4$ stability region, on a $128\times128$ grid at the limit of stability of the standard LBGK, $\omega = 1.99692$, $Re\approx 20\times 10^3$, (+, solid line), and at $\omega = 1.999$, $Re\approx 61\times 10^3$ (unstable in LBKG), ($\times$, dashed line). The region inside each line is stable up to at least $200\times 10^3$ timesteps ($t = 62.5$).}
  \label{Fig1}
\end{figure}

For the stability analysis a fixed grid size of $128\times 128$ was used. At this grid size it was found that the standard LBGK scheme becomes unstable at $\omega=1.99692$ which corresponds to ${\rm Re}\approx 20\times 10^3$. A series of simulations with the present model were then run at the same value, $\omega=1.99692$, with various values of the two free parameters $r_3$ and $r_4$. Results are presented in Fig.\ \ref{Fig1}, where points show limiting values around which many simulations with varying $r_3$ and $r_4$ were run, the points representing the limits of stable simulation. The stability domain inside these points corresponds to the successful (stable) simulations. The value of $\omega$ was then increased to $\omega=1.999$, giving ${\rm Re}\approx 61\times 10^3$, the result also being shown in Fig.\ \ref{Fig1}. Clearly an almost convex domain of stable values is found, within which all combinations of $r_3$ and $r_4$ produce a stable result. This domain is large at the standard LBGK stability limit, becoming smaller as the Reynolds number is increased into values unstable for the standard LBGK. The accuracy of solutions within such a domain are analysed in the following section.

\begin{figure}[ht]
  \begin{center}
  \begin{tabular}{cc}
	   \includegraphics[width=0.4\textwidth]{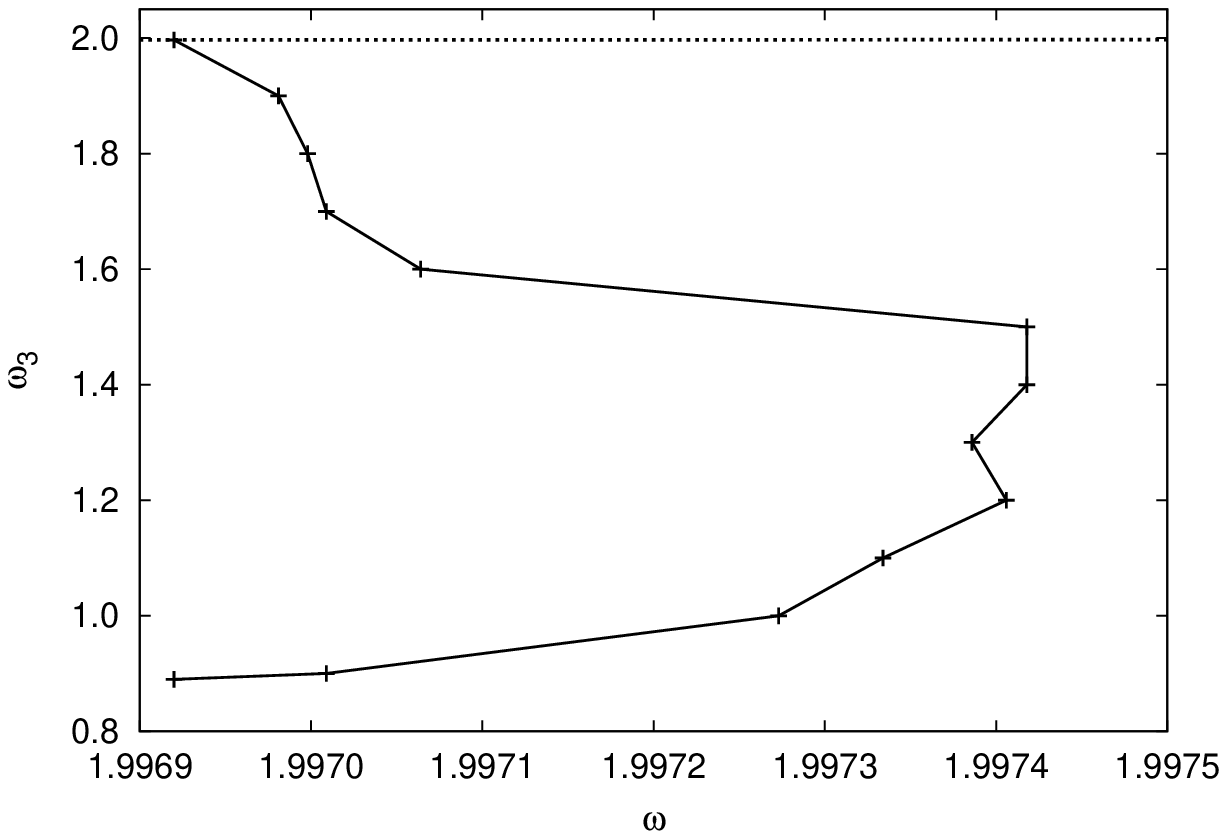}  &
       \includegraphics[width=0.4\textwidth]{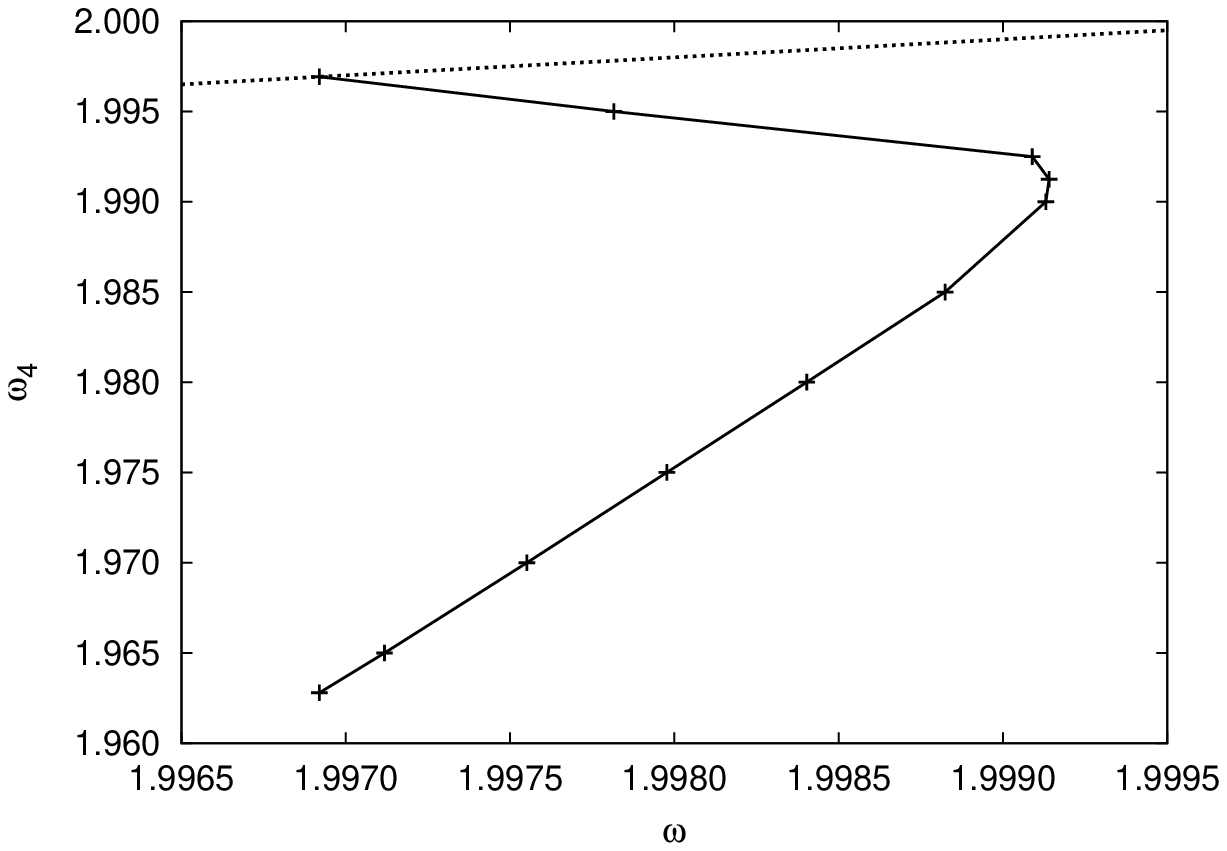}
	\end{tabular}
  \end{center}
  \caption{Stability limit of $\omega$ on a $128\times128$ grid, for (left) varying $\omega_3$ with fixed $\omega_4 = \omega$ and (right) varying $\omega_4$ with fixed $\omega_3 = \omega$. Values between the trend lines (solid lines) are stable. The dotted lines show the LBGK values. Significant increases over the standard LBGK in the stable values of $\omega$, and therefore Reynolds number, are observed.}
  \label{Fig32}
\end{figure}

Fixing $\omega_4 = \omega$, the stability limit of $\omega$ was determined over a range of values of $\omega_3$, the results being shown in Fig. \ref{Fig32}. Also shown are the results of fixing $\omega_3 = \omega$, and finding stable values of $\omega$ over a range of $\omega_4$. {In the first case the highest $\omega$ to remain stable was $\omega = 1.99742$, simulated at $\omega_3 = 1.5$, which corresponds to a Reynolds number of about ${\rm Re}\approx 24\times 10^3$. In the second case the highest $\omega$ to remain stable was $\omega = 1.99914$, simulated at $\omega_4 = 1.99125$, which corresponds to a Reynolds number of about ${\rm Re}\approx 71\times 10^3$. Varying $\omega_3$ and $\omega_4$ together gives significant further increase in stability, for example $\omega_3 = \omega_4 = 1.98$ remains stable up to $\omega = 1.999942$, corresponding to a Reynolds number of ${\rm Re}\approx 1\times 10^6$, 50 times greater than the maximum stable Reynolds number using the standard LBGK.}


The grid chosen for this set of numerical experiments is too coarse to assess the accuracy of the method.
It is well known that insufficient resolution in the present benchmark results in spurious vortices which contaminate the simulation. Many conventional numerical methods, as studied in Minion and Brown \cite{Minion97}, are shown to produce spurious vortices on $128\times 128$ grids at Reynold number of $\mathcal{O}(10^4)$.

It should be stressed that the instability of the standard LBGK was indeed triggered by these spurious vortex structures, that is, due to lack of resolution. The present model is able to sustain under these circumstances even at much higher Reynolds numbers. We therefore proceed with the accuracy study of the present model under grid refinement.

\subsection{Accuracy}

\begin{figure}[ht]
  \begin{center}
  \begin{tabular}{cc}
	   \includegraphics[width=0.4\textwidth,angle=-90]{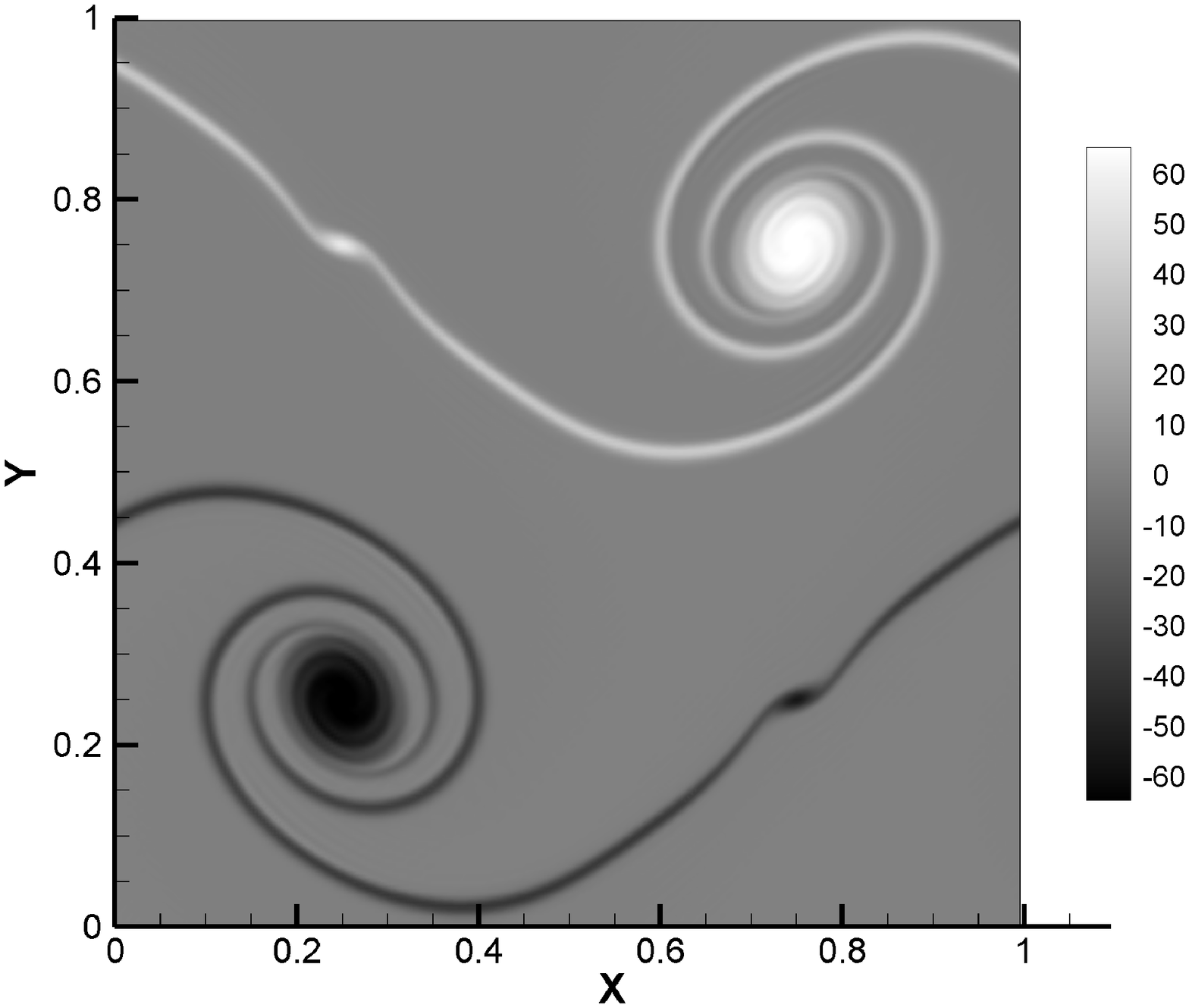}  &
        \includegraphics[width=0.4\textwidth,angle=-90]{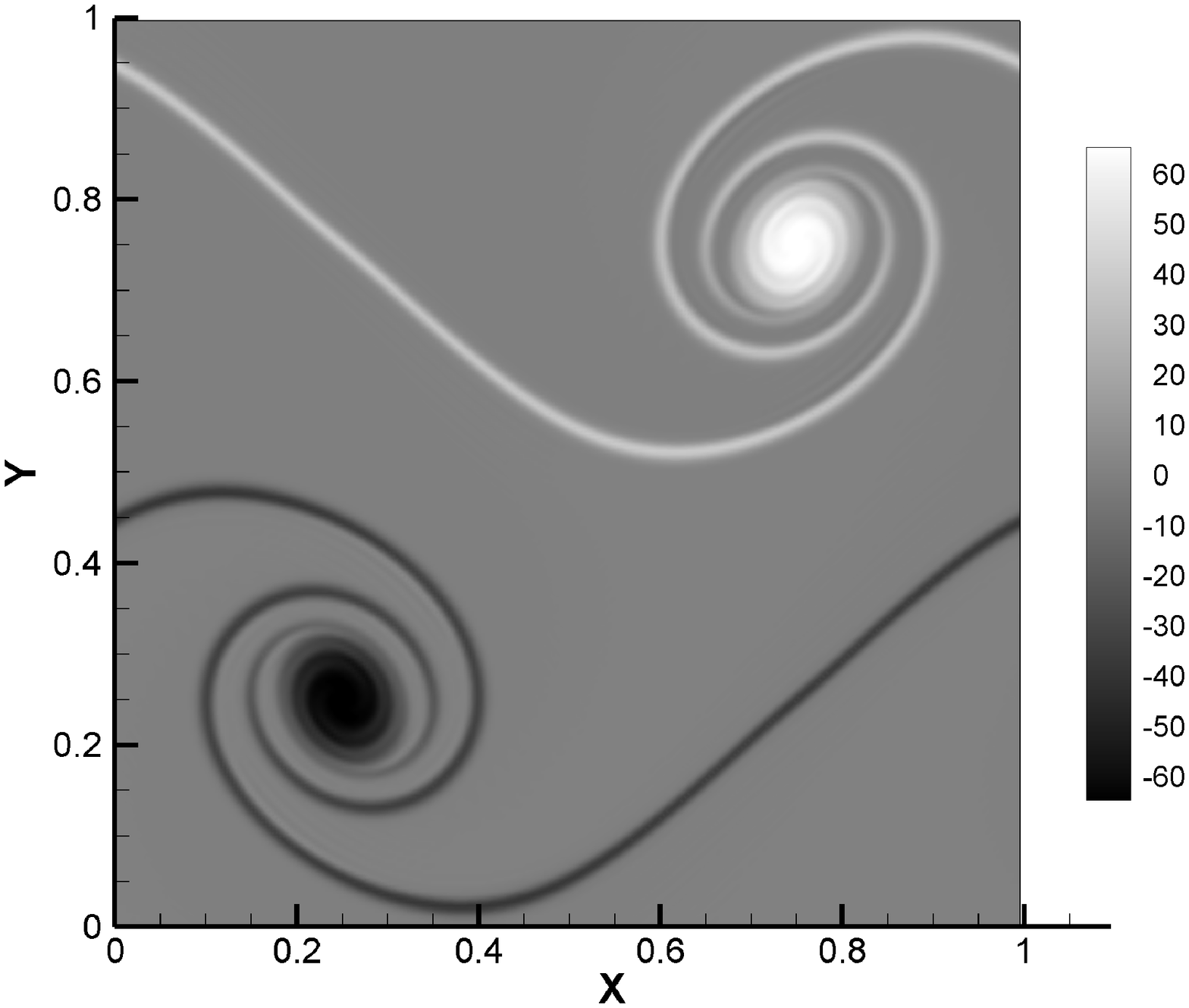}
	\end{tabular}
  \end{center}
  \caption{Vorticity field at $t = 1$, on a $256\times256$ grid with ${\rm Re}=30\times 10^3$, for the standard LBGK (left), and for the present Enhanced LBGK with $\omega_3 = \omega$, $\omega_4 = 1.99$ (right).
  The spurious vortices present for the standard LBGK are completely removed with the Enhanced LBGK.}
  \label{Fig45}
\end{figure}


In order to assess the accuracy of the present scheme, the Reynolds number was initially fixed at ${\rm Re}=30\times 10^3$, while the grid was doubled in each direction. The increase in the resolution stabilized the standard LBGK, however, the spurious vortices were still present, as shown in Fig. \ref{Fig45}. The simulation was run with the present scheme using the following sets of values: 1) $\omega_3 = 1.0, \omega_4 = \omega$, 2) $\omega_3 = \omega, \omega_4 = 1.99$, and 3) $\omega_3 = 1.0, \omega_4 = 1.9$, as shown in 
Fig. \ref{Fig45}.
It can clearly be seen that for 
$\omega_4 = 1.99$ the spurious vortices are completely removed. For $\omega_3 = 1.0, \omega_4 = 1.9$ the spurious vortices are removed even on a $128\times128$ grid. On the smaller grid the lower resolution thickens the shear layer slightly, however it is worth noting that on the larger grid the spurious vortices are completely removed for $\omega_4 = 1.99$ without any noticeable increase in the width of the shear layers, as would occur if spurious vortices were suppressed through an additional artificial viscosity (cf. e. g. Ref. \cite{Minion97}).

\begin{table}
\begin{tabular}{|l|c|}
  \hline
  Method & Minimum Resolution  \\
  \hline
  LBGK & $288\times288$  \\
  $\omega_3 = 1.5$ & $280\times280$ \\
  $\omega_4 = 1.99$ & $248\times248$ \\
  $\omega_3 = 1.8, \omega_4 = 1.95$ & $192\times192$ \\
  $\omega_3 = 1.0, \omega_4 = 1.90$ & $168\times168$ \\
  $\omega_3 = 0.5, \omega_4 = 1.80$ & $144\times144$ \\
  \hline
\end{tabular}
\caption{Minimum resolution for which spurious vortices are not observed at ${\rm Re} = 30\times 10^3$.}
\label{Table1}

\end{table}
\begin{figure}[ht]
  \begin{center}
    \includegraphics[width=0.6\textwidth]{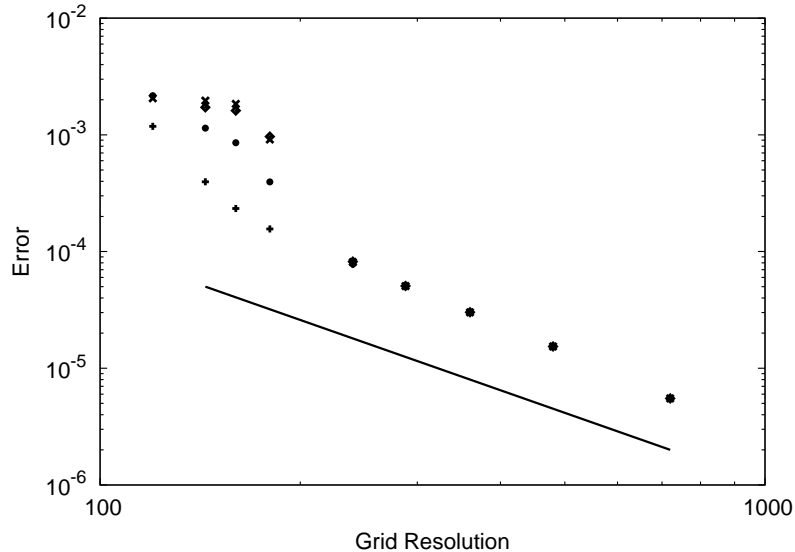}
  \end{center}
  \caption{Errors between solution on a $1440\times1440$ grid and varying grid sizes, showing convergence of the standard LBGK ($\diamond$) and the present method with 1) $\omega_3 = 1.94, \omega_4 = \omega$ ($\times$), 2) $\omega_4 = \omega, \omega_4 = 1.97$ ($\bullet$), 3) $\omega_3 = 1.7, \omega_4 = 1.6$ ($+$). In each case $p$, the rate of convergence, is 2.16. The solid line shows second order convergence, $p = 2$. (Points overlap at grid size $\ge 240\times240$.)}
  \label{Fig12}
\end{figure}

The improvement provided by the present method can be more clearly seen in Table \ref{Table1}. Here the approximate grid resolutions at which spurious vortices disappear are given for both the standard  LBGK and the various setups of the present scheme. This again shows the present method providing a clear advantage over the standard LBGK. The minimum grid resolution required to remove spurious vortices on the standard LBGK is $288\times288$, compared with only $144\times144$ using the present method. In terms of computational time, looking at grid size alone this represents an eightfold reduction to give a solution of equal accuracy, although a small overhead is incurred with this method. This also represents a fourfold reduction in required memory.

While the removal of spurious vortices provides obvious qualitative improvements in results, it is useful to compare the convergence of the current method with that of the standard LBGK. For this a solution at a high resolution, $1440\times1440$, is produced in each case at ${\rm Re} = 20\times 10^3$. The average differences in the $x$-component of velocity of this result compared with those of varying grid resolution are shown in Fig. \ref{Fig12}. All setups are seen to have approximately the expected second-order convergence, however in each case this deteriorates below a certain resolution. In agreement with the results of Table \ref{Table1}, this happens for a higher grid resolution in the standard LBGK case, and decreases to lower grid resolutions with the present scheme, the deterioration in second order convergence being due to the formation of the spurious vortices. {These results also confirm that the present scheme follows the same behaviour as the standard LBGK, the convergence rates being equal to those of LBGK.} This lends further evidence to the present scheme being an improvement on the standard LBGK, without compromising its underlying quality.

\begin{figure}[ht]
  \begin{center}
  \begin{tabular}{cc}
	   \includegraphics[width=0.4\textwidth]{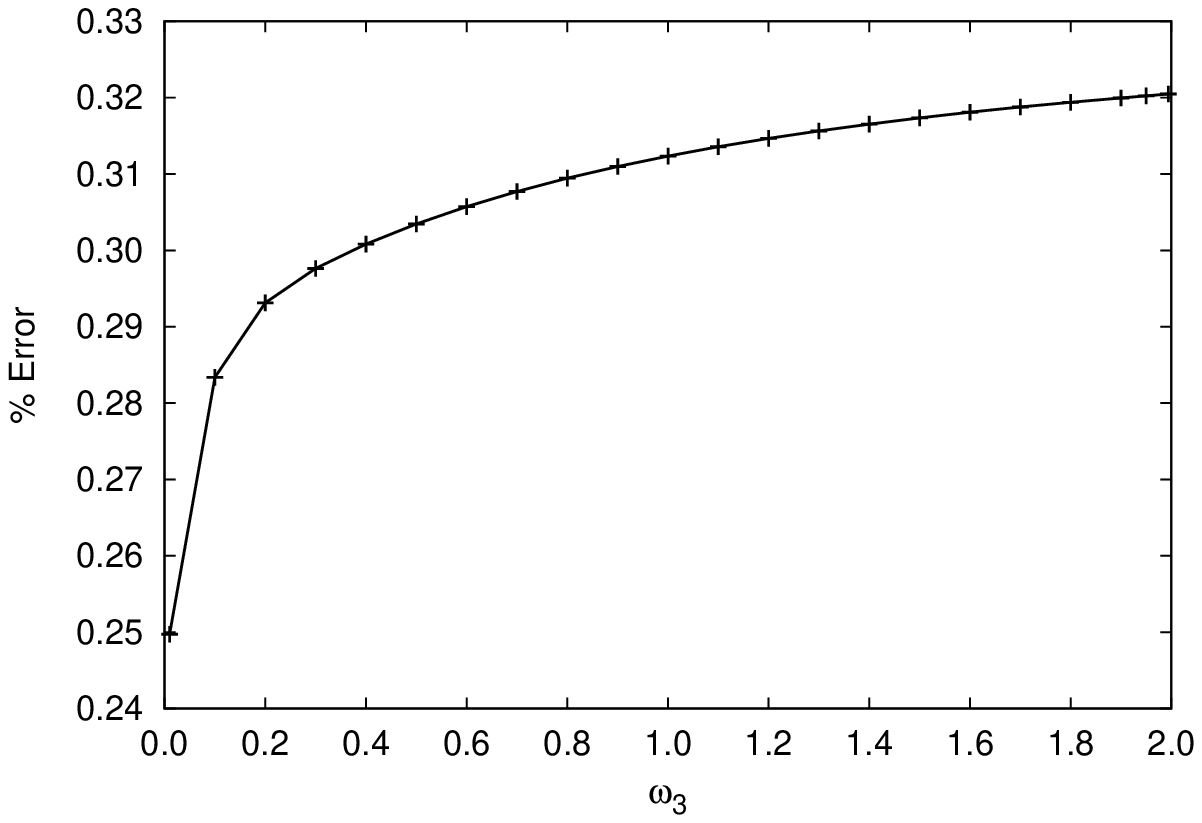}  &
       \includegraphics[width=0.4\textwidth]{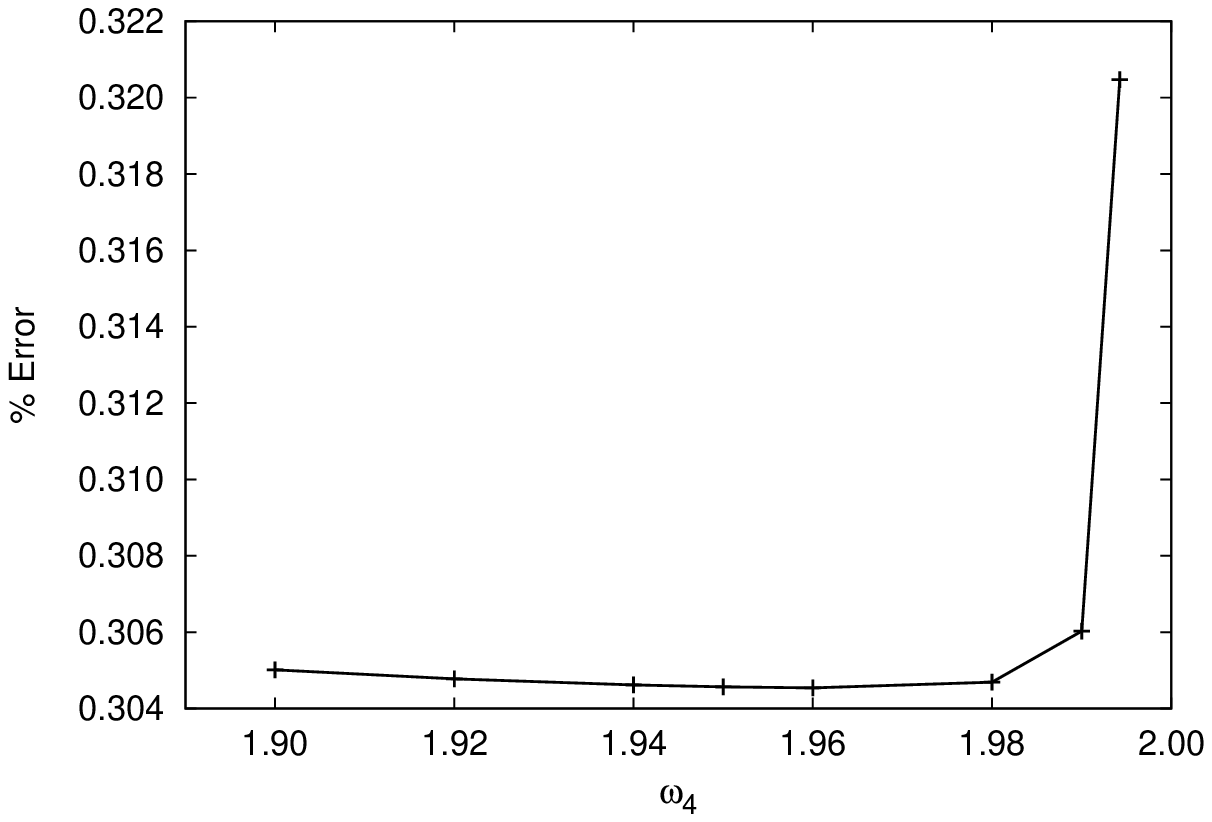}
	\end{tabular}
  \end{center}
  \caption{Errors in the $x$-component of velocity, as a percentage of average $x$-velocity, between the Enhanced LBGK on a $240\times240$ and the standard LBGK on a $1440\times1440$ grid, for $\omega_3$ at fixed $\omega_4 = \omega$ (left) and $\omega_4$ at fixed $\omega_3 = \omega$ (right).}
  \label{Fig11ab}
\end{figure}

\begin{figure}[ht]
  \begin{center}
    \includegraphics[width=0.4\textwidth]{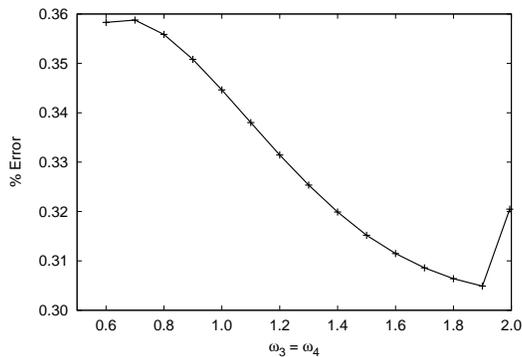}
  \end{center}
  \caption{Errors in the $x$-component of velocity, as a percentage of average $x$-velocity, between the Enhanced LBGK on a $240\times240$ and the standard LBGK on a $1440\times1440$ grid, for varying $\omega_3 = \omega_4$.}
  \label{Fig11c}
\end{figure}

The same high resolution solution is used to assess the accuracy of points within an $\omega_3$ vs $\omega_4$ domain at fixed $\omega$, as is observed in Fig. \ref{Fig1}. Here a $240\times240$ grid is used as no spurious vortices are observed at this resolution at ${\rm Re} = 20\times 10^3$, throughout the stable domain. Results are compared with the $1440\times1440$ LBGK solution. Plots of errors in $x$-velocity are made for three slices through this domain: 1) $\omega_4$ = $\omega$, $\omega_3$ varied, 2) $\omega_3$ = $\omega$, $\omega_4$ varied, and 3) $\omega_3$ = $\omega_4$ varied, with results given in Figs. \ref{Fig11ab} and \ref{Fig11c}. At this Reynolds number the average error in the standard LBGK case is $0.32 \%$. It can be seen that errors are very similar to those in the standard LBGK case, throughout the stability domain.

{For varying $\omega_3$ alone, errors are smaller than for the standard LBGK, as is also the case for varying $\omega_4$ alone. For varying $\omega_3 = \omega_4$, while for some values errors are slightly higher than for LBGK, they are of the same order. The increased stability of the present method has not affected the underlying accuracy of LBGK. In addition, a range of values of the parameters in the present model give the same quality of solution.}


\section{Conclusion}
\label{sec:conclu}

An enhanced LBGK model has been developed which demonstrates a significantly enlarged domain of stability compared with the standard LBGK. This enhancement is obtained without any significant computational overhead above the standard LBGK. In that respect the present formulation can be preferable to other methods of realization of the additional relaxation of non-conserved modes.
A large overall gain in stability is found in a benchmark simulation, with up to 50 times increase in accessible Reynolds number reported. This increase in stability is obtained without the use of artificial viscosity. A domain of relaxation parameters has been found that allow stable simulation, within which the accuracy of solution is consistent regardless of the choice of parameter values. This parameter range shrinks with increasing Reynolds number which indicates that existing approaches based on the additional relaxation times have a limitation as they do not provide unconditional stability. This study shows that the present approach is useful as it requires lower resolution grids to produce the same accuracy of solution as the standard LBGK. This enables an overall eight times reduction in computational effort as compared with the standard LBGK. The three-dimensional realization of the present model is straightforward and will be addressed in our follow-on work.

{Finally, we reiterate that the method developed herein follows the idea of arranging for different relaxation rates of central moments in a co-moving reference frame, as was first expressed by Geier et al \cite{Geier2006}. However, the present realization is different from the cascaded LB method of Ref.\ \cite{Geier2006}. While the cascaded LB method \cite{Geier2006} stabilizes the simulation with the help of artificial viscosity (which can sometimes result in various artifacts such as a re-laminarization reported in \cite{Freitas2011}), the present realization retains the accuracy of the standard LBGK, as was demonstrated in the simulations above. In that respect, the cascaded LB method can be regarded as a sub-grid model whereas the present realization is more suitable for the direct numerical simulation.}

\section*{Acknowledgments}

We gratefully acknowledge the funding from the Engineering and Physical Sciences Research Council for Grant No. EP/I000801/1
and a HEC Studentship.

\bibliography{Enhanced-LBGK-refs}

\end{document}